\documentclass[10pt,A4paper,conference]{IEEEtran}

\usepackage{amsmath}
\usepackage{upref}
\usepackage{amsfonts}
\usepackage{graphicx}
\usepackage{latexsym}
\usepackage{amssymb}

\newcommand{\mathset}[1]{\left\{#1\right\}}
\newcommand{\abs}[1]{\left|#1\right|}

\newcommand{\parenv}[1]{\left( #1 \right)}

\newtheorem{defn}{Definition}
\newtheorem{theorem}{Theorem}

\newtheorem{cor}{Corollary}

\DeclareMathOperator{\pr}{Pr}

\begin{document}

\title{On the Asymptotic Performance of Iterative Decoders for
Product Codes
}

\author{\authorblockN{Moshe Schwartz}
\authorblockA{University of California San Diego\\
9500 Gilman Drive, Mail Code 0407 \\
La Jolla, CA 92093-0407, U.S.A.\\
{\tt moosh@everest.ucsd.edu}}
\and
\authorblockN{Paul H.~Siegel}
\authorblockA{University of California San Diego\\
9500 Gilman Drive, Mail Code 0401\\
La Jolla, CA 92093-0401, U.S.A.\\
{\tt psiegel@ucsd.edu}}
\and
\authorblockN{Alexander Vardy}
\authorblockA{University of California San Diego\\
9500 Gilman Drive, Mail Code 0407\\
La Jolla, CA 92093-0407, U.S.A.\\
{\tt vardy@kilimanjaro.ucsd.edu}}
}

\maketitle

\begin{abstract}
We consider hard-decision iterative decoders for product codes
over the erasure channel, which employ repeated
rounds of decoding rows and columns alternatingly. We derive the exact
asymptotic probability of decoding failure as a function of the
error-correction capabilities of the row and column codes, the number
of decoding rounds, and the channel erasure probability.
We examine both the case of codes capable of correcting a constant amount
of errors, and the case of codes capable of correcting
a constant fraction of their length.
\end{abstract}

\section{Introduction}

One of the simplest methods of combining two codes is the product construction.
Let $C_1$ and $C_2$ be $[n_1,k_1,d_1]$ and $[n_2,k_2,d_2]$ linear codes
respectively. Then, the set of $n_1\times n_2$ arrays whose columns are
codewords of $C_1$ and whose rows are codewords of $C_2$, is the product
code $C_1\otimes C_2$ with parameters $[n_1 n_2, k_1 k_2, d_1 d_2]$.

Product codes are useful in a variety of applications (for references see
\cite{MacSlo78}). They may be found in the ubiquitous 
CD standard IEC-908 and CD-ROM standard ECMA-130 (for details see
\cite{Imm91}),
as well as the DVD standard ({\tt www.dvdforum.org}).
Their rectangular shape makes them especially
appealing to two-dimensional error-control applications (for references
see \cite{Far92}).

The fact that product codes retain entire codewords of their constituent codes
makes it tempting to use an iterative decoder in the following fashion.
First, each of the columns is decoded using a decoder for $C_1$. The resulting
(partially) decoded array is then used for a new round of decoding in which
each row is decoded using a decoder for $C_2$. This process may be carried
for any number of rounds deemed necessary, alternatingly decoding rows
and columns. It is therefore a natural question to ask what is the
decoding-failure probability of such a scheme.

This probability is obviously a function of the amount of errors correctable
by the columns and row codes, the number of decoding rounds, and the
channel error probability. The channel need not be binary, and we assume that
it introduces an error
in a certain position independently of other positions, and with the same
distribution. We use the row and column decoders as black boxes with the only
assumption being that they do not misdecode, i.e., they either correct all the
errors or do nothing. This is the case when we take the $q$-ary erasure
channel.

In this work we calculate the asymptotic decoding-failure probability.
This is a rare case where we can precisely predict the performance
of iterative decoding analytically for a specific code rather than
a random ensemble.
We manage this by recasting the problem to a graph-theoretic setting in which
the channel is thought of as producing random bi-partite graphs. By doing
so, we can rely on well-known mechanisms for checking properties of
random graphs.

The paper is organized as follows. In Section \ref{sec:background} we give
some necessary background on random graphs. We continue to
Section \ref{sec:fixed}, in which we examine row and column codes which
can correct a constant number of errors. This will be referred to as the
{\em constant error correction case}. In Section \ref{sec:good}
we handle the case of row and column codes capable of correcting a constant
fraction of their length, which we will call the {\em linear error
correction case}. We conclude in Section \ref{sec:conclusion} with a
discussion of the results.


\section{Background}
\label{sec:background}

Let us examine the case where we have a product code with codewords of
size $n\times n$, and the column code and row codes are each capable of
correcting $t$ errors. After passing through
the channel, a received word may contain errors. We can represent these
errors as a bi-partite graph $G=(V_L,V_R,E)$ with a vertex in
$V_L=\mathset{1,\dots,n}$ for each row, and a vertex in
$V_R=\mathset{n+1,\dots,2n}$ for each column. An edge $(i,j)$ is in $E$
if and only if position $(i,j-n)$ is in error.

Let $n_L$ and $n_R$ be positive integers, and let $0\leq p\leq 1$. The
{\em random bi-partite graph,} $G(n_L,n_R,p)$, is a probability space over the
set of bi-partite graphs on the vertex set $V=V_L\cup V_R$, where
$\abs{V_L}=n_L$, $\abs{V_R}=n_R$, $V_L\cap V_R=\emptyset$,
and with
$$\pr[(i,j)\in G]=\begin{cases}
p & i\in V_L, j\in V_R \\
0 & \text{otherwise,}
\end{cases}$$
with these mutually independent.
Hence, we can think of
our channel as producing a random bi-partite graph $G(n,n,p)$.

A {\em round} of decoding consists of an attempt to decode either all the
rows, or all the columns. When viewed in the bi-partite graph representation,
a decoding round consists of going over either the vertices of $V_L$ or
the vertices of $V_R$, and for each vertex with degree less or equal to $t$,
removing all of its adjacent edges.
Successive rounds alternate between rows and columns.
We assume, w.l.o.g., that the last round is always performed on the rows.
We denote the number of rounds as $r$, where $r$ is a constant.

Given a bi-partite graph, $G$, representing the transmission errors, we
say that it is {\em $(r,t)$-decodable} if a decoder which can correct up to $t$
errors in each row and column, corrects all the errors after $r$ rounds.
In other words, after $r$ rounds of edge removals as described above, no
edges remain.
Our aim is to analyze the asymptotic probability of decoding failure.

Throughout this paper we follow the notation of \cite{AloSpe00}.
Given two functions, $f(n)$ and $g(n)$, we say that $f(n)\ll g(n)$
if $f(n)=o(g(n))$.
Let $G$ be a graph, and let $A$ be a graph-theoretic property. If $G$
has property $A$, we denote it by $G\models A$. In our case, $A$ is the
property that the graph is $(r,t)$-decodable. As mentioned in \cite{AloSpe00},
many graph-theoretic properties exhibit a threshold behavior as follows.

\begin{defn}
$r(n)$ is called a threshold function for a graph theoretic property $A$
if

\noindent$\bullet$
When $p(n)\ll r(n)$, $\lim_{n\rightarrow\infty}\pr[G(n,n,p(n))\models A]=0$.

\noindent$\bullet$
When $p(n)\gg r(n)$, $\lim_{n\rightarrow\infty}\pr[G(n,n,p(n))\models A]=1$.
\end{defn}

Finally, given some event $A$ whose probability depends on some parameter $n$,
we say that $A$ occurs {\em almost always} if
$\lim_{n\rightarrow\infty}\pr[A]=1$.


\section{The Constant Error Correction Case}
\label{sec:fixed}

In this section we handle the case where both the column and row codes
are capable of correcting a constant $t$ number of errors. This is done
by first noting that a bi-partite graph representing the transmission errors
is decodable if and only if it does not contain a certain subgraph which
we call an {\em $(r,t)$-undecodable configuration}. We then continue,
using the theory of random graphs, to analyze the probability that the
random bi-partite graph contains this undecodable configuration.
We need the following definitions first.

\begin{defn}
Let $G=(V,E)$ be a graph, and let $v\in V$ be
a vertex of the graph. We denote by $N_i(v)$ the set of vertices
of $G$ which are reachable from $v$ by a path (not necessarily simple)
of length exactly $i$.
\end{defn}

We note that under this definition, a vertex of degree at least 1 is
its own neighbor at distance 2, since we can take a path going over an
outgoing edge, and returning by the same edge. 
In fact, such a vertex is its own neighbor for any even distance.
We denote the degree of vertex $v\in V$ as $d(v)$.

\begin{defn}
Let $G=(V_L,V_R,E)$ be a bi-partite graph. We say that $G$ is an
{\em $(r,t)$-undecodable configuration} if there exists $v\in V_L$
such that all the following hold:
\begin{itemize}
\item
$\bigcup_{i=0}^r N_i(v)=V_L\cup V_R$.
\item
For all $0\leq i\leq r-1$ and $v'\in N_i(v)$, $d(v')\geq t+1$.
\end{itemize}
We call $v$ the {\em root} of $G$.
\end{defn}

Note that by the first requirement, the graph must be connected.
Hence, the sets $N_i(v)$ are not disjoint, and both
$$N_0(v)\subseteq N_2(v)\subseteq N_4(v) \subseteq \dots\\$$
and
$$N_1(v)\subseteq N_3(v)\subseteq N_5(v) \subseteq \dots$$
hold. Furthermore, if $v'\in N_i(v)$ for some $i\geq 0$, then
its immediate neighbors are all in $N_{i+1}(v)$.

\begin{defn}
Let $G=(V_L,V_R,E)$ be a bi-partite graph, and let $H=(V'_L,V'_R,E')$ be
another bi-partite graph. We say that $H$ is an {\em ordered
bi-partite subgraph} of $G$ if there exist injective functions
$f_L:V'_L\rightarrow V_L$ and $f_R:V'_R\rightarrow V_R$ such that if
$(v_1,v_2)\in E'$, then $(f_L(v_1),f_R(v_2))\in E$.
\end{defn}

The following theorem is the basis for our analysis.

\begin{theorem}
A bi-partite graph $G=(V_L,V_R,E)$ is $(r,t)$-decodable if and only if
it does not have an $(r,t)$-undecodable configuration
as an ordered bi-partite subgraph.
\end{theorem}
\begin{proof}
In the first direction, let us assume that $G$ contains an $(r,t)$-undecodable
graph $H$ as an ordered bi-partite subgraph. Let $v$ be the root of $H$.
In the first round of decoding $G$, the vertices corresponding to $N_{r-1}(v)$
are not decoded since they have degree of at least $t+1$. In the following
round, if we take any vertex $v'\in N_{r-2}(v)$, all of its neighbors
are in $N_{r-1}(v)$ so they were not corrected in the first round.
Since $v'$ has degree at least $t+1$, it follows that the vertices of
$N_{r-2}(v)$ are not corrected in the second round. Continuing in the same
manner, after $r$ rounds, the sole vertex of $N_0(v)=\mathset{v}$ is not
corrected, so $G$ is not $(r,t)$-decodable.

In the other direction, let $G$ be a graph which is not $(r,t)$-decodable.
Hence, after $r$ rounds of decoding, there exists a vertex $v\in V_L$ which
was not corrected. We now show that $v$ is the root of an
$(r,t)$-undecodable configuration $H=(V'_L,V'_R,E')$
which is an ordered bi-partite subgraph
of $G$. We start by obviously defining $N_0(v)=\mathset{v}$ and taking
$V'_L=\mathset{v}$, $V'_R=\emptyset$, and $E'=\emptyset$. Now, since $v$
was not decoded at the end of round $r$, round $r-1$ ended with
$v$ having at least $t+1$ undecoded neighbors. We denote this set of neighbors
as $N_1(v)$. We also add these neighbors to $V'_R$ and the appropriate edges
to $E'$. Take some $v'\in N_1(v)$. Since $v'$ was not decoded at the
end of round $r-1$, round $r-2$ ended with $v'$ having at least $t+1$ undecoded
neighbors. Going over all possible $v'\in V'_R$, and taking the union of the
undecoded neighbors we get $N_2(v)$. Note that $N_0(v)\subseteq N_2(v)$.
We add $N_2(v)$ to $V'_L$ and the appropriate edges to $E'$. Continuing
in the same manner we get an $(r,t)$-undecodable configuration
as the theorem states.
\end{proof}

By the previous theorem, the question of undecodability becomes a purely
graph-theoretic question.
For the asymptotic analysis we need the following definitions and
probabilistic tools.

\begin{theorem}[\cite{AloSpe00}]
\label{th:expect0}
Let $X$ be a non-negative integral-valued random variable.
If $E[X]=o(1)$, then $X=0$ almost always.
\end{theorem}

\begin{proof}
Trivial.
\end{proof}

For the rest of this section, let $X=X_1+\dots+X_m$ where
$X_i$ is the indicator random variable for event $A_i$. For indices
$i$, $j$, we write $i\sim j$ if $i\neq j$ and events $A_i$ and $A_j$
are not independent. We define
$$\Delta=\sum_{i\sim j} \pr[A_i\wedge A_j].$$

\begin{theorem}[Corollary 4.3.4, \cite{AloSpe00}]
\label{th:delta}
If $E[X]\rightarrow\infty$ and $\Delta=o(E[X]^2)$, then $X>0$
almost always.
\end{theorem}

\begin{defn}[\cite{AloSpe00}]
Let $H$ be a graph with $v$ vertices and $e$ edges. We call $\rho=e/v$ the
{\em density} of $H$. We call $H$ {\em balanced} if every subgraph $H'$ has
$\rho(H')\leq \rho(H)$. We call $H$ {\em strictly balanced} if every
proper subgraph $H'$ has $\rho(H')<\rho(H)$.
\end{defn}

The following is an adaptation of Theorem 4.4.2, \cite{AloSpe00}, to
bi-partite random graphs.

\begin{theorem}
\label{th:balancedthreshold}
Let $H$ be a balanced bi-partite graph with $v$ vertices and $e$ edges.
Let $G(n,n,p)$ be a random bi-partite graph, and let $A$ be the event
that $H$ is an ordered bi-partite subgraph of $G$. Then $p=n^{-v/e}$
is the threshold function for $A$.
\end{theorem}
\begin{proof}
Let $H=(V'_L,V'_R,E')$ be a balanced bi-partite graph. Denote
$v_L=\abs{V'_L}$, and $v_R=\abs{V'_R}$, so $v=v_L+v_R$. Let $G=(V_L,V_R,E)$
be a random bi-partite graph. Let $S$ be a $v$-subset of the vertices of $G$
such that $\abs{S\cap V_L}=v_L$ and $\abs{S\cap V_R}=v_R$. Let $A_S$ be the
event that the subgraph of $G$ induced by $S$ contains $H$ as an
ordered bi-partite subgraph. Then obviously,
$$p^e\leq \pr[A_S] \leq v_L! v_R! p^e.$$
Let $X_S$ be the indicator random variable for $A_S$ and
$$X=\sum_S X_S.$$
By linearity of expectation,
$$E[X]=\sum_S E[X_S]=\binom{n}{v_L}\binom{n}{v_R}\Pr[A_S]=\Theta(n^v p^e).$$
If $p(n)\ll n^{-v/e}$ then $E[X]=o(1)$, so by Theorem \ref{th:expect0},
$X=0$ almost always.

Now assume $p(n)\gg n^{-v/e}$ so that $E[X]\rightarrow\infty$,
and consider $\Delta$ of Theorem \ref{th:delta}.
$$\Delta=\sum_{S\sim T} \pr[A_S\wedge A_T]=
\sum_S \pr[A_S]\sum_{T\sim S}\pr[A_T | A_S].$$
Here, two $v$-sets
$S$ and $T$ satisfy
$S\sim T$ if and only if $S\neq T$ and they share some edges,
i.e., $\abs{S\cap T\cap V_L}\geq 1$ and $\abs{S\cap T\cap V_R}\geq 1$.
Let $S$ be fixed, so
$$\sum_{T\sim S}\pr[A_T | A_S]=
\sum_{i=2}^{v-1}\sum_{\substack{
\abs{S\cap T}=i \\
\abs{S\cap T\cap V_L}\geq 1\\
\abs{S\cap T\cap V_R}\geq 1}} \pr[A_T | A_S].$$
For each $i$ there are $O(n^{v-i})$ choices of $T$. Fix $S$ and $T$, and
consider $\pr[A_T | A_S]$. There are $O(1)$ possible copies of $H$ on $T$.
Since $H$ is balanced, each has at most $\frac{ie}{v}$ edges with both vertices
in $S$, hence at least $e-\frac{ie}{v}$ other edges. Therefore,
$$\pr[A_T | A_S]=O(p^{e-\frac{ie}{v}}),$$
and
\begin{multline*}
\sum_{T\sim S}\pr[A_T | A_S]=\sum_{i=2}^{v-1}O(n^{v-i}p^{e-\frac{ie}{v}})\\
=\sum_{i=2}^{v-1}O((n^v p^e)^{1-\frac{i}{v}})=
\sum_{i=2}^{v-1}o(n^v p^e)=o(E[X]),
\end{multline*}
since $p(n)\gg n^{-v/e}$. We have already seen that
$\pr[A_S]=O(p^e)$, and there are $O(n^v)$ choices for $S$, so
\begin{align*}
\Delta & =\sum_S \pr[A_S]\sum_{T\sim S}\pr[A_T | A_S] \\
&=O(n^v p^e)o(E[X])=o(E[X]^2).
\end{align*}
By Theorem \ref{th:delta}, $X>0$ almost always.
\end{proof}

Let us start by examining one specific type of an $(r,t)$-undecodable
configuration.
We define an {\em exact $(r,t)$-undecodable tree} as an
$(r,t)$-undecodable configuration without cycles in which each vertex
at distance at most $r-1$ from the root has degree $t+1$ exactly.
It is easy to see that such a graph is strictly balanced. We denote
the number of edges in such a tree as $e_T(r,t)$. This number is easily
seen to be:
\begin{equation}
\label{eq:et}
e_T(r,t)=\begin{cases}
2r & t=1\\
(t+1)\frac{t^r-1}{t-1} & t\geq 2.
\end{cases}
\end{equation}
Since this is a tree, obviously the number of vertices $v_T(r,t)$ is
exactly $e_T(r,t)+1$. By Theorem \ref{th:balancedthreshold},
the threshold function for the existence of an exact $(r,t)$-undecodable
tree in $G(n,n,p)$ is
\begin{equation}
\label{eq:tree}
p=n^{-\parenv{1+\frac{1}{e_T(r,t)}}}.
\end{equation}

Another case is when the $(r,t)$-undecodable tree is not exact, i.e.,
the configuration is a tree, but
some vertices at distance at most $r-1$ from the root have a degree which
is strictly more than $t+1$. However, in such a case, the existence of
a non-exact $(r,t)$-undecodable tree implies the existence of an
exact $(r,t)$-undecodable tree (simply trim the excess edges and vertices).

Thus we are left with the case of $(r,t)$-undecodable configurations which
are not trees at all. Such configurations must contain cycles. It is also
easy to see that such configurations must contain a simple cycle with at
most $2r$ edges.

If we take a graph of a simple cycle with $e$ edges, it also has $e$ vertices.
This graph is also strictly balanced. It follows that
the threshold function for
the existence of such a cycle is $p=n^{-1}$. Hence, when
\eqref{eq:tree} holds, or when
$$p\ll n^{-\parenv{1+\frac{1}{e_T(r,t)}}},$$
there are almost always no simple cycles of length
at most $2r$. This is because each length almost always does not appear,
and there
are $O(1)$ such lengths which interest us, so a simple union bound suffices.
Thus, there are almost
always no $(r,t)$-undecodable configurations with cycles in $G(n,n,p)$
under these conditions.

\begin{cor}
The threshold function for the existence of an $(r,t)$-undecodable
configuration in $G(n,n,p)$, for some fixed $r$ and $t$, is
$$p=n^{-\parenv{1+\frac{1}{e_T(r,t)}}},$$
where $e_T(r,t)$ is given by \eqref{eq:et}.
\end{cor}

Now that we have established a threshold behavior for the existence of
$(r,t)$-undecodable configurations, we are left with the case where
$$p=c\cdot n^{-\parenv{1+\frac{1}{e_T(r,t)}}},$$
for some constant $c>0$. We know that in this case,
the question of the existence of an
$(r,t)$-undecodable configuration in $G(n,n,p)$ reduces to the question
of the existence of an exact $(r,t)$-tree in $G(n,n,p)$.

For the following analysis we need the Janson inequality \cite{Jan90}.
Let $\Omega$ be a finite universal set, and let $R$ be a random
subset of $\Omega$ given by
$$\pr[r\in R]=p_r,$$
where these events are mutually independent. Let $B_i$, $i\in I$ be
subsets of $\Omega$, where $I$ is a finite index set. Let $A_i$ be the
event that $B_i\subseteq R$. Let $X_i$ be the indicator variable for
$A_i$, and let $X=\sum_{i\in I}X_i$. We denote the complementary
event to $A_i$ as $\overline{A_i}$. We set
$$M=\prod_{i\in I}\pr[\overline{A_i}].$$

\begin{theorem}[The Janson Inequality, \cite{Jan90}]
\label{th:janson}
Let $A_i$, $i\in I$, $\Delta$, and $M$, be as above, and assume that
$\pr[A_i]\leq\epsilon$ for all $i\in I$. Then
$$M\leq \pr[\wedge_{i\in I}\overline{A_i}]
\leq M e^{\frac{1}{1-\epsilon}\frac{\Delta}{2}}.$$
\end{theorem}

We can now continue by adapting Theorem 10.1.1, \cite{AloSpe00},
to bi-partite graphs.

\begin{theorem}
Let $H$ be a strictly-balanced bi-partite graph with $v$ vertices,
$e$ edges, and $a$ automorphisms. Let $c>0$ be some constant.
We denote by $A$ the event
that $G$ does not have $H$ as an ordered bi-partite subgraph. Then,
when $p=c\cdot n^{-v/e}$ we have,
$$\lim_{n\rightarrow\infty}\pr[G(n,n,p)\models A]=exp[-c^e/a].$$
\end{theorem}
\begin{proof}
Let $H=(V'_L,V'_R,E')$ be a strictly-balanced bi-partite graph. Denote
$v_L=\abs{V'_L}$, and $v_R=\abs{V'_R}$, so $v=v_L+v_R$. Let
$B_i$, $1\leq i\leq \binom{n}{v_L}\binom{n}{v_R}v_L! v_R!/a$,
range over the edge sets of possible placements of $H$ as an ordered
bi-partite subgraph of $G$. Let $A_i$ be the event that
$B_i\subseteq G(n,n,p)$.

We use Janson's inequality from Theorem \ref{th:janson}.
For all $i$, $\pr[A_i]=p^e$, so
$$\lim_{n\rightarrow\infty}M=
\lim_{n\rightarrow\infty}(1-p^e)^{\binom{n}{v_L}\binom{n}{v_R}v_L! v_R!/a}
=exp[-c^e/a],$$
since $p=c\cdot n^{-v/e}$. We turn to handle
$$\Delta=\sum_{i\sim j}\pr[A_i \wedge A_j].$$
When $i\sim j$, let $k$ denote
the number of vertices in the intersection of the two
placements of $H$. Obviously $2\leq k\leq v$. Let $f_k$ denote the
maximal number of edges in the intersection when $i\sim j$ and there
are $k$ vertices in the intersection. For $k=v$ we obviously have
$f_v<e$ since $i\neq j$. When $2\leq k\leq v-1$, since $H$
is strictly-balanced and $B_i\cap B_j$ is a subgraph of $H$,
$$\frac{f_k}{k} < \frac{e}{v}.$$
There are $O(n^{2v-k})$ choices of $i$ and $j$ which intersect in $k$
vertices. Hence, for each such $i$ and $j$,
$$\pr[A_i\wedge A_j]=p^{\abs{B_i\cup B_j}}=p^{2e-\abs{B_i\cap B_j}}
\leq p^{2e-f_k},$$
and then
$$\Delta=\sum_{k=2}^v O(n^{2v-k})O(n^{-\frac{v}{e}(2e-f_k)}).$$
But
$$2v-k-\frac{v}{e}(2e-f_k)=\frac{v f_k}{e}-k < 0,$$
so $\Delta=o(1)$. Janson's inequalities become a sandwich, so
$$\lim_{n\rightarrow\infty}\pr[\wedge_i\overline{A_i}]
=\lim_{n\rightarrow\infty}M=exp[-c^e/a].$$
\end{proof}

Fortunately, an exact $(r,t)$-tree is strictly balanced, and has
the following number of automorphisms:
\begin{equation}
\label{eq:automorphism}
a_T(r,t)=\begin{cases}
2 & t=1 \\
(t+1)!(t!)^{(t+1)\frac{t^{r-1}-1}{t-1}} & t\geq 2.
\end{cases}
\end{equation}
Thus we get the following corollary:
\begin{cor}
The probability that $G(n,n,p)$ is $(r,t)$-decodable when
$p=c\cdot n^{-\parenv{1+\frac{1}{e_T(r,t)}}}$, is asymptotically
$exp[-c^{e_T(r,t)}/a_T(r,t)]$, where $e_T(r,t)$ is given by \eqref{eq:et}
and $a_T(r,t)$ is given by \eqref{eq:automorphism}.
\end{cor}

Finally, we summarize the case of constant error correction in Table
\ref{tab:constant}.

\begin{table}[h]
\caption{The asymptotic probability of decoder failure with $r$ rounds of
decoding, a constant $t$ of decodable errors in each row and column,
and erasure probability $p$}
\label{tab:constant}
\begin{center}
\begin{tabular}{|c|l|}
\hline
$p(n)$ & Decodability \\
\hline\hline
$p\ll n^{-\parenv{1+\frac{1}{e_T(r,t)}}}$ &
Almost always decodable \\
\hline
$p=c\cdot n^{-\parenv{1+\frac{1}{e_T(r,t)}}}$ &
\begin{minipage}{1.4in}
Decodable with probability \\ $exp[-c^{e_T(r,t)}/a_T(r,t)]$
\end{minipage}
\\
\hline
$p\gg n^{-\parenv{1+\frac{1}{e_T(r,t)}}}$ &
Almost always undecodable \\
\hline
\end{tabular}
\end{center}
\end{table}


\section{The Linear Error Correction Case}
\label{sec:good}

We now turn to the case of linear error correction capabilities. This
case appears to be much simpler than the previous one. Let
$F=\mathset{C_1,C_2,\dots}$ be an infinite family of codes of ever
increasing length, and let us denote the length of $C_i$ by $n_i$.
We now require that $C_i$ is capable of correcting $\alpha n_i$ errors,
where $0<\alpha<1$ is some constant. We note that ``good'' codes
also fall into this category.

Just like before, in this section we consider a product code having codewords
of size $n\times n$. For convenience, the row code and the column code are
each capable of correcting $\alpha n$ errors, where $0<\alpha<1$ is a constant.
We denote by $p$ the erasure probability of the channel.

The main tool for our analysis is the well known Chernoff bound.

\begin{theorem}[The Chernoff Bound]
\label{th:chernoff}
Let $X=\sum_{i=1}^n X_i$ be the sum of $n$ independent random variables,
each in $[0,\delta]$. Let $\epsilon\in(0,1]$, and $\mu>0$, be fixed. Then,
\begin{itemize}
\item
If $E[X_i]\leq\mu$ for all $i$, then $\pr[X\geq (1+\epsilon)\mu n] <
exp[-\epsilon^2\mu n/ (3\delta)]$.
\item
If $E[X_i]\geq\mu$ for all $i$, then $\pr[X\leq (1-\epsilon)\mu n] <
exp[-\epsilon^2\mu n/ (2\delta)]$.
\end{itemize}
\end{theorem}

\noindent
We can now state the main result.

\begin{theorem}
Let $0\leq p\leq 1$ be fixed. Then,
\begin{itemize}
\item
If $p<\alpha$ then using only one round of decoding, i.e., only the row code,
the decoder successfully decodes any received word almost always.
\item
If $p>\alpha$ then no matter how many rounds of decoding are used, the
decoder fails to decode any received word almost always.
\end{itemize}
\end{theorem}
\begin{proof}
Assume $p<\alpha$. Choose some fixed $0<\epsilon<\alpha/p-1$. Using the
Chernoff bound of Theorem \ref{th:chernoff}, any given row
contains less than $\alpha n$ errors with probability tending to 1
exponentially fast. Hence, by a simple union bound, the probability of the
event that there is some undecodable row in the first round, tends to 0
exponentially fast. Hence, the first round of decoding successfully
corrects all errors almost always.

If $p>\alpha$, choose some fixed $0<\epsilon<1-\alpha/p$. Just like the 
previous case, the probability that any given row or column contain
less than $\alpha n$ errors tends to 0 exponentially fast by the
Chernoff bound. Hence, given a received word, by a simple union bound, the
probability that there is any row or column which is decodable tends
to zero exponentially fast. This means, that almost always the decoder fails
on all the rows and all the columns. Hence, no matter how many rounds are
used, the decoding process fails almost always.
\end{proof}


\section{Conclusion}
\label{sec:conclusion}

We analyzed the asymptotic probability of decoding failure of iterative
decoders for product codes. Our analysis is limited to the case of
hard-decision decoding over the erasure channel. We examined both the case
of codes capable of correcting a constant number of errors in each row and
column, and the case of codes capable of correcting a constant fraction
of the length of each row and column.

In the case of constant error correction, the asymptotic probability shows
a threshold behavior. As shown in Table \ref{tab:constant}, when the
erasure probability decays faster than the threshold function, we can
correctly decode every received word almost always. Conversely, when it
decays slower, we almost always fail to decode any received word. When
the erasure probability decays just like the threshold function up to
a multiplication by a constant, we have an exact expression for the
probability of decoder failure. It should be noted, that it is
beneficial to invest
in more rounds of decoding, and codes which correct more errors, since both
$e_T(r,t)$ and $a_T(r,t)$ are increasing functions and $a_T(r,t)$ grows
faster than $e_T(r,t)$. This means that higher values of $r$ and $t$ give
threshold functions closer to $n^{-1}$, and higher probability of
successfully decoding when the erasure probability is at the threshold.

The other case, of linear error correction capabilities, is perhaps more
curious. Again we have a sharp threshold behavior, and in this case, it
is constant. However, above this threshold, we almost always have too many
errors for each row or column to correct and we fail, no matter how many
rounds of decoding we do. Below this threshold, we almost always eliminate
all the errors after the first round of decoding, so one round of decoding
is enough. In that case, we do not get to use the column code at all, and
the redundancy invested in it -- is simply redundant. So it appears in that
case, that it is better to use just the row code instead of the product code.

\section*{Acknowledgment}

This work was supported by a research grant from Applied Micro Circuits
Corporation, San Diego, CA.
The authors would also like to thank Roy Schwartz for some helpful discussions.

\bibliography{allbib}

\end{document}